# Principle of Entangled-Photon Thermometry for Ultrafast Laser Processing

**Denys Moskal[1]*, Jiří Martan[1], Šimon Kos[2], Miroslav Holeček[1], Milan Honner[1], Vladislav Lang[1]**

[1]New Technologies Research Centre (NTC), University of West Bohemia in Pilsen, Univerzitní 8, 301 00 Pilsen, Czech Republic

[2]Department of Physics and NTIS – European Centre of Excellence, University of West Bohemia in Pilsen, Univerzitní 8, 301 00 Pilsen, Czech Republic

*Correspondence: moskal@zcu.ntc.cz

## Abstract

A quantum-enhanced approach for fast temperature diagnostics in ultrashort laser surface processing is introduced. The goal is to overcome the current limitations of the measuring methods, like influence of plasma emission, changes of emissivity during ablation, or time consuming pump-probe setup. The core principle of the described method is exploiting polarization anisotropy in entangled photon pairs. In this method, the idler photon interacts with the laser-affected material surface, while its entangled counterpart, the signal photon, is detected in a remote, undisturbed optical arm. The strong correlation of the photons' polarizations after the interaction that provides the idler photon measurement is crucial for the method. Temperature-dependent changes in the complex refractive index modulate the reflectance of *p*- and *s*-polarizations on the idler path, which in turn alter the coincidence-resolved polarization statistics of the signal photons. Using a Qiskit-based quantum model incorporating experimentally measured pump-probe reflectometry data, we show that this entanglement-assisted scheme enables remote reconstruction of rapid thermal dynamics during femtosecond laser ablation. This approach is not experimental, however experimental data embedded in the simulation algorithm were taken from literature sources to model a realistic material response. Due to the binary and highly discrete nature of single-photon events, classical sliding-window analysis exhibits strong shot noise or temporal inertia. To overcome this limitation, we apply a multi-layer perceptron (MLP) regression network that extracts implicit anisotropy information from the photon bitstream. The neural-network approach yields an order-of-magnitude reduction in temperature noise, enhances signal-to-noise ratio (SNR), and improves temporal accuracy to the nanosecond scale. These results demonstrate that entangled-photon polarization anisotropy, combined with machine-learning analysis, offers a promising route toward undisturbed, background-rejected, high-speed temperature diagnostics in laser–matter interaction studies.

## 1. Introduction

Laser surface texturing (LST) using ultrashort pulses initiates a high temperature process, such as spallation or phase explosion. Part of laser energy is absorbed under the surface in the latent form of a liquid phase or later chemical changes, like oxidation of upper layers [1]–[3]. During high frequency laser pulsing, such accumulation processes become significant and are able to change the quality and efficiency of LST. To evaluate the thermal changes between laser pulses it is possible to perform thermal measurements with fast IR-sensors in wave range 1-10 μm and it gives the possibility to detect phase changes or even re-structuration of surface in real time [4]–[6]. The problem is in the correct evaluation of absolute value of temperature, because the emitted IR-thermal radiation intensity is affected by optical properties of the laser processed material, plasma glow and non-homogeneous temperature distribution. On the other hand, laser surface processing significantly modifies the optical properties of the material. This leads to a coupled problem in which temperature measurements depend on the optical properties, while the optical properties themselves are affected by temperature changes through the emissivity [3], [7]. The existing methods, which are in use to bypass this principal limitation, are following the dynamics of IR-thermal radiation changes and using some plateaus as benchmarks which correspond to fixed temperature points, like melting or solidification transitions. Another method is two-color measurement in which the relative changes of the IR-thermal signal on two (or more) sensors with different optical high pass and low pass filters are in use for temperature changes evaluation [6], [8], [9]. Such a method is based on the spectral distribution of IR described by Planck's law, followed by Wien's displacement law [10]. A higher temperature increases emission at shorter wavelengths and alters the intensity ratio between the shorter and longer wavelength bands defined by the IR detector filters [11], [12]. But even this two-color pyrometry method is affected by optical changes to the irradiated surface, like phase, chemical or structural changes [6], [11], [12]. An alternative is in a fast measurement of optical changes by using an additional beam of light with reflection from laser ablated spot and evaluation of optical properties changes as indicator of temperature changes [3], [13]. The method of the reflected beam intensity detection cannot be used for the direct temperature evaluation, because reflection changes in the laser spot interaction area not only by the temperature itself, but also by the surface deformation and its structure changes during ablation. An alternative approach is the application of fast polarization-resolved thermometry using a pulsed pump–probe ellipsometry method. The temperature changes are then evaluated not from the intensity of the reflected beam, but from the analysis of the changes in the p- and s-polarization components of the reflected beam. Ultrafast pump–probe ellipsometry is capable of evaluating temporal changes in optical properties and determining temperature variations during the laser ablation process, with a time resolution equal to the pulse duration—that is, on the picosecond or even femtosecond timescale [14], [15]. The limitations of this method include the need for consequent sample movement to a fresh surface position for each time-step measurement. Additional challenges arise from changes in the reflection angle due to thermal modification of the sample surface, as well as from the separation of the reflected beam from ablation plasma emission, thermal radiation, and scattered light.

The improvement of the existing optical methods of the fast temperature measurement can be achieved by application of quantum effects, like entangled states with polarization and coincidence detection. Such a method of 'entangled twin-photon' ellipsometer [16]–[19] has the separate arms of the ellipsometer and the light beams traverse them independently in different directions. This allows any desired polarization manipulation in the untouched signal arm of the entangled twin-photon ellipsometer, independently from changing of incidence angle on the idler arm [16], [19]. In this paper, we discuss a possible application of the same twin-photon ellipsometer in the context of time-resolved temperature measurement. Unlike the mentioned twin-photon ellipsometer the principle, which is described in this paper, involved a fixed polarization basis in combination of quantum entangled photons and polarization-resolved thermometry.

## 2. Coincidence and temperature measurements with two-photon entangled state

### 2.1. Detection of temperature changes by polarization anisotropy

In this article, the temperature measurement process is investigated through a proposed simulation. Although it is not a direct experiment, the simulation algorithm incorporates experimental data from literature sources to model a realistic material response. A nonlinear crystal can generate pairs of entangled photons through the process known as spontaneous parametric down-conversion (SPDC) [16], [20]. In this process, an incoming high-energy pump photon interacts with the nonlinear medium and, with a small probability, “splits” into two lower-energy photons commonly called the *signal* and *idler*. Energy and momentum conservation ensure strict correlations between these two photons, while the nonlinear interaction preserves the quantum coherence necessary for entanglement. Under appropriate phase-matching conditions, determined by the crystal orientation, refractive indices, and pump wavelength, the emitted photon pair emerges in a well-defined pure quantum state.

When the crossed crystals are operated in the type-I SPDC configuration, both the signal and idler photons are produced with the same polarization. This leads naturally to a Bell-like entangled state in which the joint polarization of the pair is perfectly correlated, even though the polarization of each photon individually remains undefined until a measurement is performed. A typical state produced in this geometry can be written, for example, as:

$$| \Psi\rangle = \frac{1}{\sqrt{2}}\left(|H\rangle_{sg}|H\rangle_{id} + |V\rangle_{sg}|V\rangle_{id}\right) \tag{1}$$

where $| H\rangle$ and $| V\rangle$ denote horizontal and vertical polarization components, and the subscripts $sg$ and $id$ refer to the signal and idler photons. This expression captures the essential feature of polarization entanglement: although neither photon has a definite polarization on its own, any measurement performed on one instantly determines the polarization outcome of the other, regardless of their spatial separation.

Physically, the entanglement arises from the quantum indistinguishability of the possible down-conversion pathways inside the crystal. The pump photon may convert into either a pair of horizontally polarized photons or a pair of vertically polarized photons. In the absence of any environmental information that could distinguish between these alternatives, the system is forced by quantum mechanics into a coherent superposition of both processes. This superposition is what gives rise to the correlated outcomes typical of Bell states (Eq. 1) [21], [22]. In an optically isotropic system, the probabilities of measuring H and V polarization are equal for both the signal and idler photons.

Thus, SPDC provides a robust and experimentally accessible source of entangled photon pairs in which both photons always share the same polarization. This makes type-I SPDC an ideal platform for studies of polarization correlations, Bell-inequality violations, quantum communication protocols, and relevant to the present work – indirect detection experiments that rely on polarization-correlated absorption or filtering applied to one branch of the entangled pair. When the idler photon is absorbed in the basis state $H$ or $V$, the entangled photon collapses into the corresponding basis state $H$ or $V$, respectively:

$$| \Psi_s^{(H)} \rangle = | H \rangle \tag{2}$$

$$| \Psi_s^{(V)} \rangle = | V \rangle \tag{3}$$

Also in case of the idler photon reflection, the signal photon collapses into the basis state with the same polarization as that of the reflected idler photon. In other words, the idler photon is absorbed or reflected either in state $| H \rangle$ or $| V \rangle$, it is dependent on optical properties of the reflected surface on the idler arm beam path. When the idler photon undergoes polarization-dependent loss or measurement, the conditional polarization statistics of the entangled signal photon reflect the same preference, and the joint detection statistics exhibit corresponding correlations [23]. The signal photon remains in the polarization state given by Eq. (1), and its $H/V$ statistics are correlated with the reflectance of the coupled idler photon through coincidence measurements. Such entangled-state correlations have nothing in common with the classical signal propagation and can be exploited in the idler arm to detect rapid optical changes, for example those induced by temperature variations. This fast and remote detection of dynamic optical properties is particularly suitable for the analysis of ultrashort laser pulse material ablation, where a simple bucket photodetector can be used on the idler side instead of a complex optical system. The bucket detector does not resolve the polarization state of the idler photon and serves solely as a trigger; however, a fixed polarization filter can be introduced at the detector to enhance filtering and isolate a selected polarization component. All polarization-resolved analysis is then performed remotely in the "clean" signal arm, where the detected data can be processed at high speed using an oscilloscope or FPGA module (Fig. 1). It should be noted that the proposed method does not provide a direct measurement of temperature; instead, it relies on calibration for a given material, enabling the evaluation of temperature-dependent processes such as heat accumulation or phase transitions.

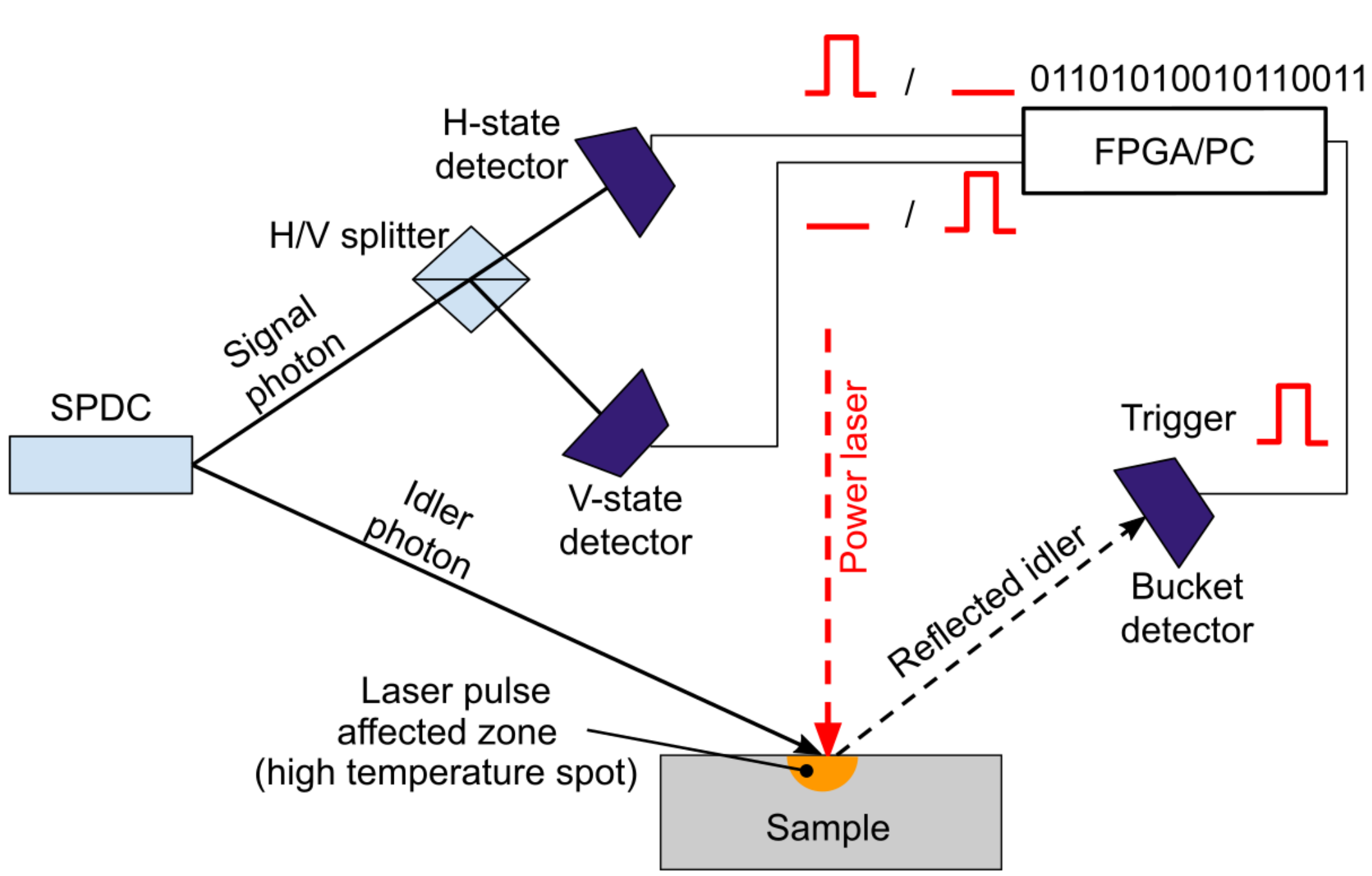

Fig. 1. Temperature measurement using photon entanglement with a signal-photon detector in the “clear-arm” and an idler-photon trigger. The photon interactions with the material act as a quantum measurement with changing polarization statistics.

In the modeling of such “clear-arm” remote temperature measurement of the femtosecond laser pulses surface heat accumulation, the process of entangled photon generation and detection is programmed with Qiskit quantum simulation toolkit [24], [25]. The entangled-photon process is modeled using Qiskit to ensure a physically consistent simulation of entanglement generation, correlation, and measurement. Although classical correlated bit generation could reproduce similar statistics, the quantum-circuit approach provides a more realistic and extensible framework aligned with the underlying experimental implementation. The temperature affected absorption of idler photons is simulated with approximation of rapid surface temperature increasing during ten picoseconds and correspondingly fast changes of optical properties [14] (adapted Fig. 2.). Such a short heating time is explained by dynamic energy exchanges between electrons and photons under two-temperature model [26].

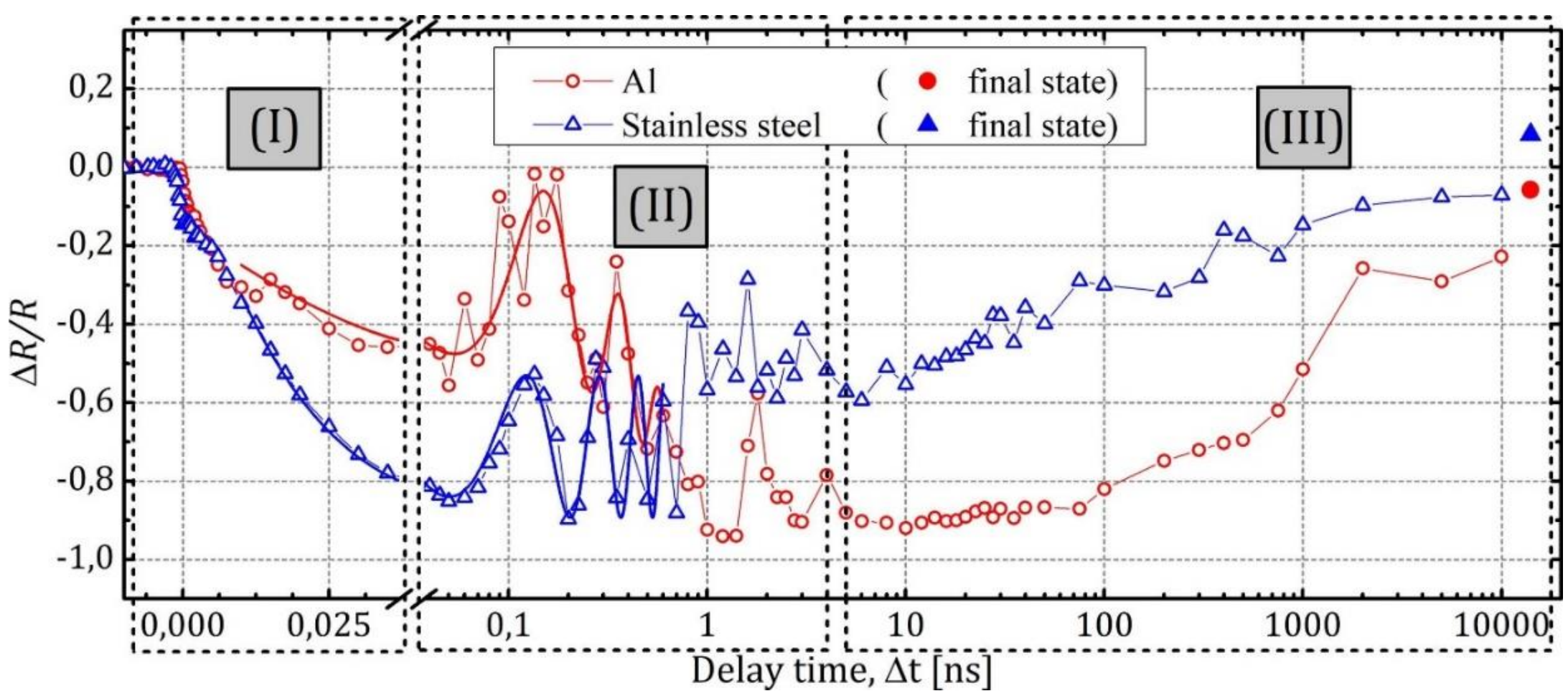

Fig. 2. Time-resolved measurements from pump-probe reflectometry (PPR) of Al and AISI304 during laser surface ablation (for experimental details see [14]).

After an ultrashort laser pulse, the surface temperature rapidly increases. The longer process of surface cooling is coming afterwards and can take several microseconds. In a subsequent nanosecond-long interval after the laser pulse, several physical processes are activated, such as spallation and phase explosion, which initiate optical wave propagation. Then latent heat keeps the cooling process longer with other corresponding processes of surface reformation [1], [4], [5], [13], [14]. At the same time period, the ablation plume is spread out above the laser processed surface. It is difficult to separate the IR thermal radiation from the laser-processed surface, as it contains several components: plasma emission, near-infrared thermal radiation from the ablated material, and infrared thermal radiation from the post-ablated surface. The influence of the mentioned effects will be suppressed by detection of a reflected beam with idler photons, which are entangled with the signal photons on the remote beam path. In our Qiskit model, the reflection of the idler photons is simulated using the wavelength-dependent real and imaginary components of the refractive index.

A rigorous derivation linking reflectivity to coincidence statistics is beyond the scope of this work. Instead, we use a phenomenological approach in which the effect is captured via reflection of idler photons entangled with the remote signal arm. In the Qiskit model, this reflection is described using time-dependent complex refractive index components n(t) and k(t), obtained by fitting pump–probe reflectance data from J. Winter et al. [15] and interpolated to the 810 nm idler wavelength. The model therefore relies on experimentally validated optical parameters and they were derived from pump–probe experimental data and interpolated to match the 810 nm idler wavelength [14]:

$$n(t) = \begin{cases} n_0, \ t < 0 \\ n_\infty + [n_0(1 - s_n H_s(t)) - n_\infty] exp\left(-\frac{t}{\tau_n}\right), \ t \geq 0 \end{cases}$$
$$k(t) = \begin{cases} k_0, \ t < 0 \\ k_\infty + [k_0(1 - s_k H_s(t)) - k_\infty] exp\left(-\frac{t}{\tau_n}\right), \ t \geq 0 \end{cases}, \quad (4)$$

where $n(t)$ and $k(t)$ are the real and imaginary parts of the time-dependent complex refractive index; $n_0$ and $k_0$ are the baseline values at negative delay (before the pump pulse); $n_\infty$ and $k_\infty$ are the late-time plateau values after relaxation; $s_n$ and $s_k$ are small prompt fractional drops at $t \gtrsim 0$ (dimensionless); $H_s(t)$ is a smooth approximation of the Heaviside function used to distinguish between the pre-pulse and laser-on regimes; $\tau_n$ and $\tau_k$ are the exponential relaxation time constants for $n$ and $k$ (in the picosecond range); and $\tau_s$ is a sub-picosecond step time constant used within the smooth Heaviside function. The reflection coefficient for both p- and s- polarization components will be changed dynamically after laser pulse application (Fig. 3). As discussed above, a rapid temperature increase leads to a drop in reflectance followed by a relatively long cooling period, after which the reflectance returns to its initial value. This general behavior is used for a functional description of reflectance changes during laser surface processing: a fast temperature rise associated with a decrease in reflectance, followed by a slow cooling stage with gradual reflectance recovery on time scales of up to 10 μs. Then according to changes of $n(t)$ and $k(t)$ it is possible to determine reflectance for both $p$- and $s$- components in the reflected beam (Fig. 3, blue and red lines correspondingly).

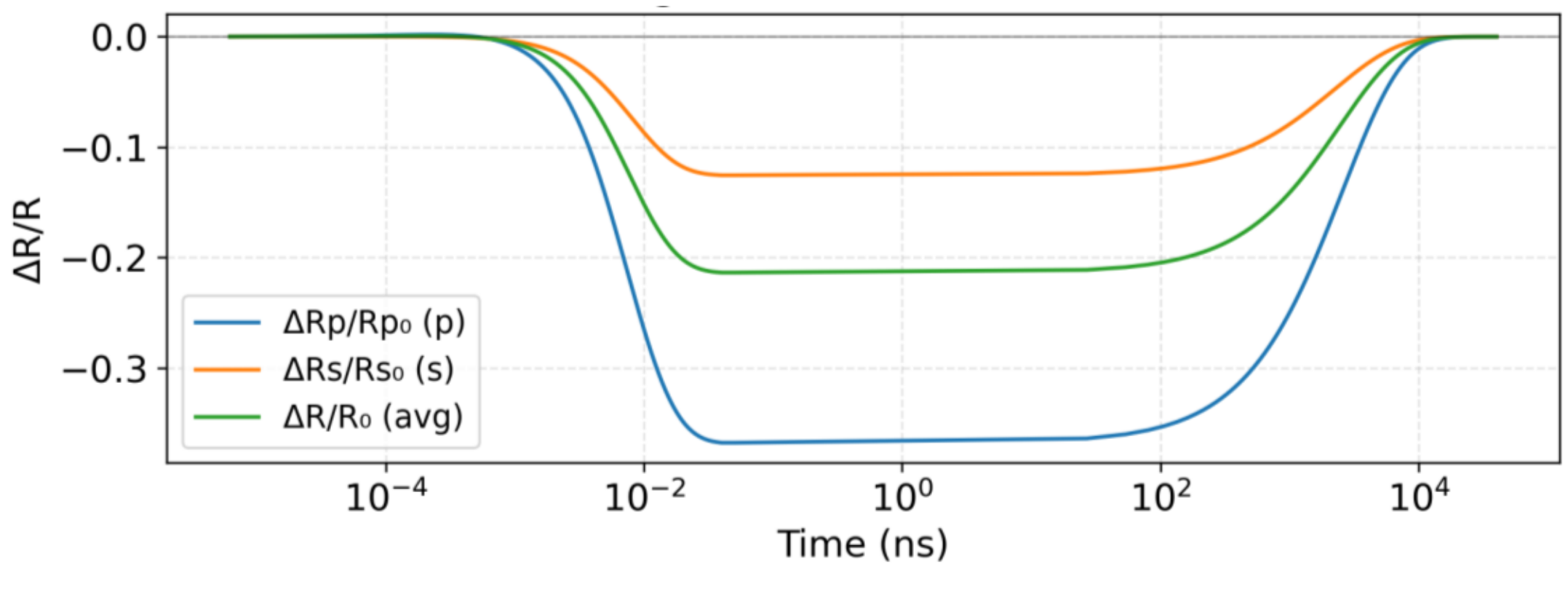

Fig. 3. Reflection dynamic changes after laser pulse application on stainless steel (based on experimental results of Winter et al [14], approximation for 810 nm).

Such reflectance changes are directly affected by temperature changes on the laser processed surface. In this logic the baseline values of refractive index $n_0$ and $k_0$ describe the initial state of the used material (corresponds to baseline values of refractive index $n_0$ and $k_0$ in Eq. 4). The plateau in the refractive index can be attributed to the existence of a liquid phase and corresponds to the melting point of stainless steel, $T_m = 1510$ °C. In any case, the exact temperature influence on the reflection coefficient can be calibrated for both polarization components, enabling the recognition of fast temperature changes via the polarization anisotropy factor:

$$A(t) = \left(R_p(t) - R_s(t)\right)/\left(R_p(t) + R_s(t)\right) \quad (5)$$

where $R_p(t)$ and $R_s(t)$ are evaluated from coincidence photon counts. There is no need to resolve the polarization of the idler photons; only their registration times, recorded by a fast single-photon bucket detector, are required. The polarization anisotropy defined by Eq. (5) represents the normalized difference in the registered coincidence-event intensities for photons reflected with $p$- and $s$-polarization, respectively. In simple terms, it is an indicator of the difference between the $p$- and $s$-polarized components in the reflected beam ($p$ can be marked as $V$ and $s$ can be marked as $H$). An alternative and simpler way to rapidly evaluate polarization anisotropy is to consider the ratio $R_p(t)/R_s(t)$. This implies that it is sufficient to detect the relative coincidence count rates of $p$- and $s$-polarized photons in the remote signal arm in order to evaluate dynamic temperature changes. Conversely, idler photons that do not produce coincidence events with the signal arm are excluded from the evaluation protocol. The quantum circuit diagram of the described remote temperature evaluation is presented in Fig. 4.

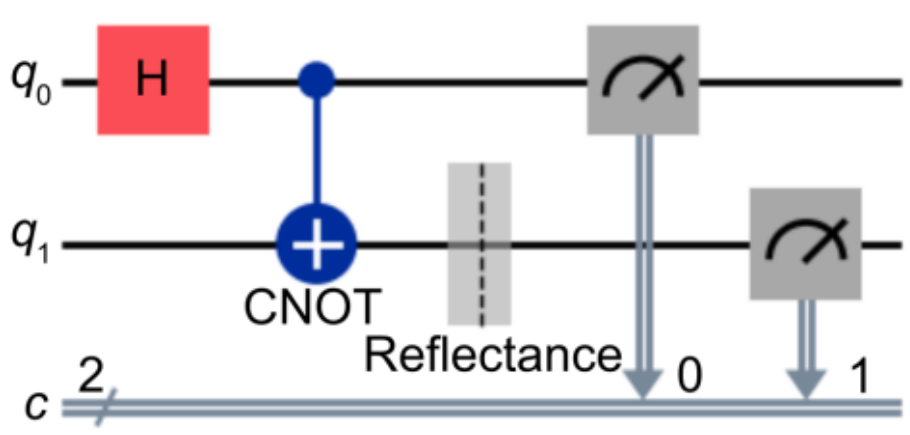

Fig. 4. Quantum diagram with two entangled arms and remote polarization detection on the signal side during reflectance on the idler side.

There are two inputs in this scheme: the first qubit ($q_0$) represents the signal photon and is generated simultaneously with the idler photon ($q_1$). In the first step, a Hadamard gate is applied to the signal qubit $q_0$, preparing it in the superposition state $|+\rangle = (|0\rangle + |1\rangle)/\sqrt{2}$. The subsequent CNOT gate (controlled-NOT) flips the state of the idler qubit conditioned on the state of the signal qubit. Due to the superposition of the signal qubit, the application of the CNOT gate creates an entangled state, namely the Bell state described by Eq. 1. When the idler beam meets a metallic plate (marked as “Reflectance”), then part of them are reflected to the bucket detector and on the signal side just coincidence $H$- or $V$-state photons are detected (red markers in Fig. 5). In the simulation, each photon detection event is modeled with a characteristic decay time of ~ 50 ps and superimposed on a white noise background [27], [28]. There are no reflectance changes in this example, just a fixed stainless steel plate with stable temperature. It means the rate between coincidences of photons in $H$-state and photons in $V$-state corresponds to initial temperature of the reflected material and can be used as a calibration point.

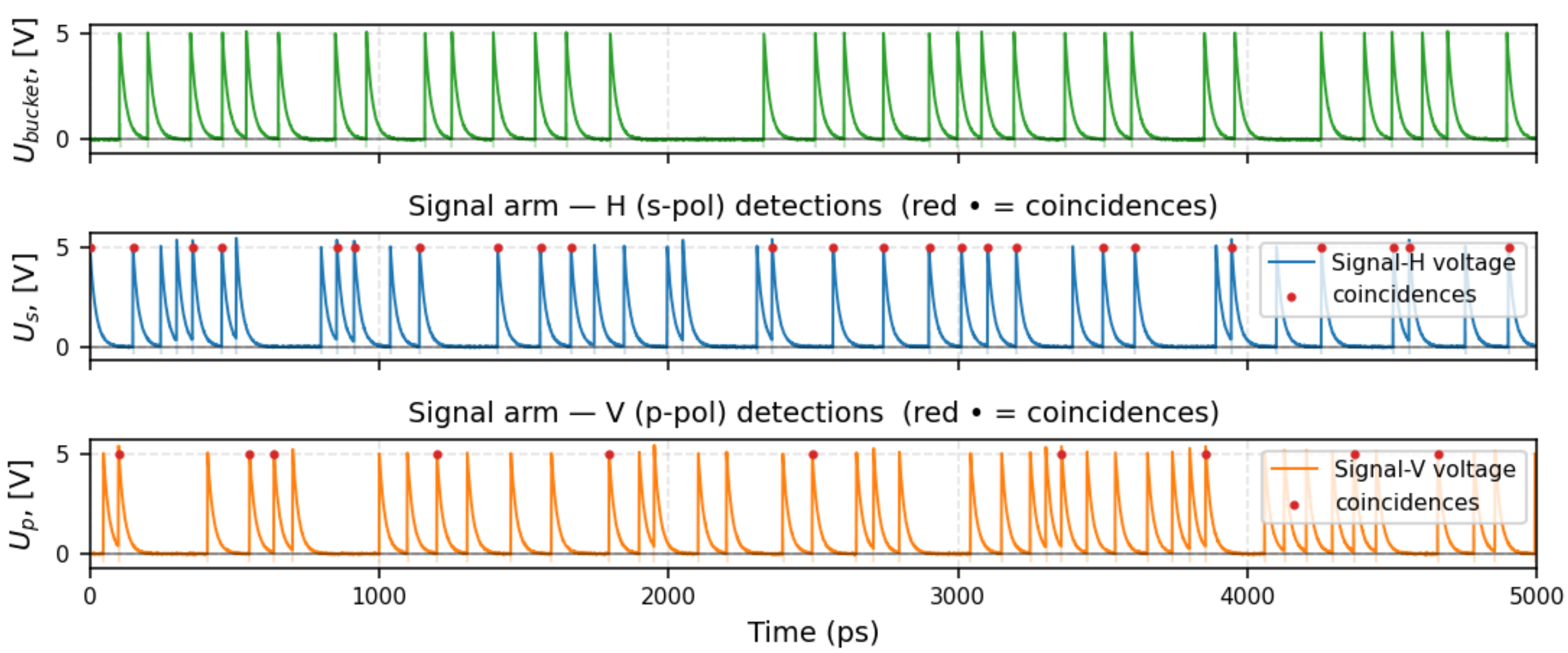

Fig. 5. Binary vector forming on triggering of signal photons in $H$- or $V$- state by reflected idler photons events.

The formation of a binary coincidence vector from $H$- and $V$-polarized signal-photon detections triggered by reflected idler photons (Fig. 5) serves as the basis for the subsequent temperature evaluation. In this approach, the polarization anisotropy of coincidence events is used directly as a temperature-sensitive observable of the reflecting surface, without reconstructing the absolute values of the refractive index $n$ and extinction coefficient $k$. The data shown correspond to a stable surface temperature and serve as a reference baseline for subsequent measurements under dynamic heating conditions. Unlike conventional ellipsometry, which requires angle-resolved measurements to evaluate $n$ and $k$, the present method is designed for temperature diagnostics based on relative polarization-dependent coincidence statistics. The incidence angle remains fixed during the measurement process.

**2.2. Detection of dynamic temperature changes of the reflecting surface**

The laser dynamic temperature changes can be involved as a function of time (Eq. 4), according to experimental results (Fig. 2 and its functional view in Fig. 3) [14]. In this case the quantity of the remotely detected signal photons will be changed according to the changes in refraction index components. The changes in $R_s$ and $R_p$ will be used for extraction of real process temperature. A quantitative analysis of the relative reflectivity change in the laser affected by the entanglement photons analysis is shown in Fig. 6. For each event the isolated evaluation of $R_p$ or $R_s$ coefficients is impossible due to incompatibility of the detection of both the $H$- and $V$-polarization state simultaneously. For the relevant evaluation of the reflection coefficient of both components, an evaluation window needs to be used. A longer window gives a smoother result, but it has higher inertia. The shorter variant gives a higher noise. For the 50 ps event of a single photon detection the evaluation window size can be taken equal up to hundreds of nanoseconds. For better resolution, it is necessary to use faster detectors or to divide the measurements into several channels. Such multichannel variant is not a problem for the remote analysis of the entanglement photon pairs, where the interacted arm works just as a fast trigger.

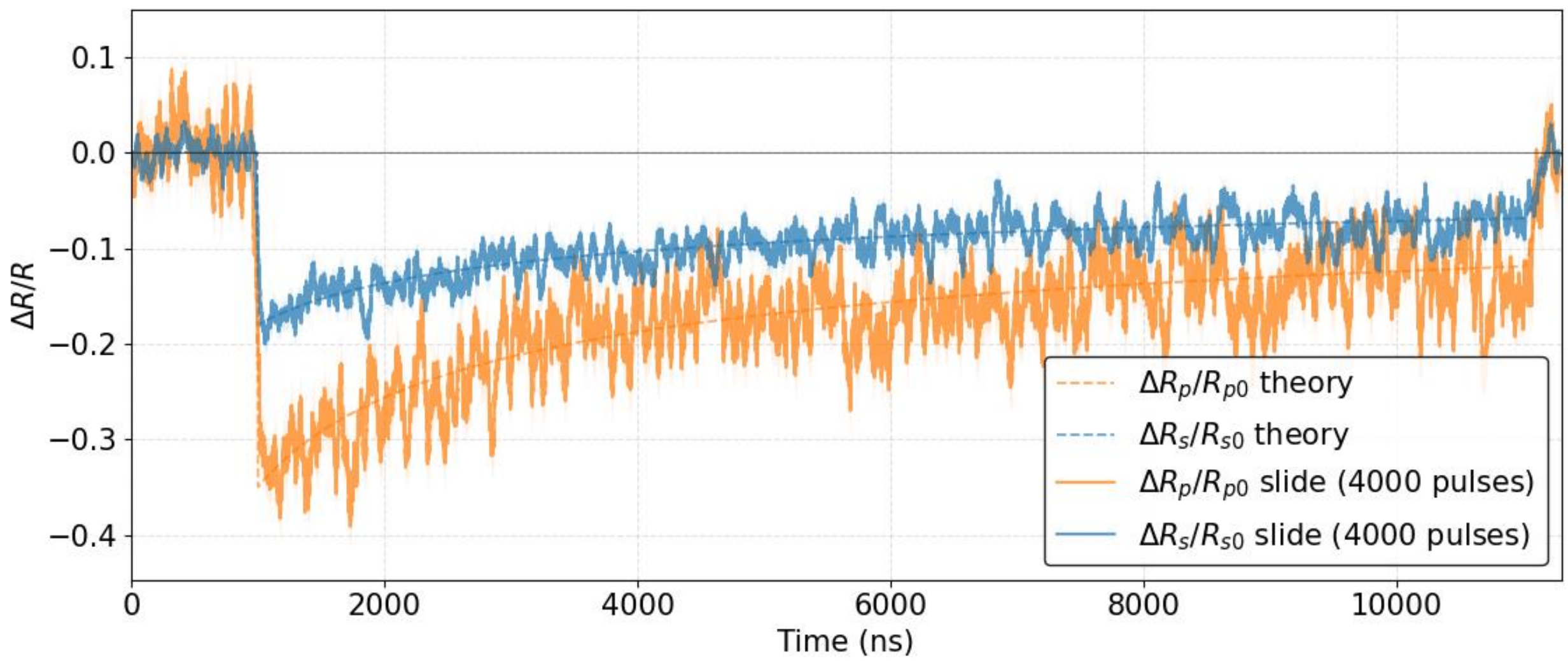

Fig. 6. Remote detection of reflection coefficients changes by entanglement photon pairs during ultrashort laser surface interaction: dashed lines - approximation of reflection coefficients from experimental data (Winter at el. [14]); solid lines - reflection coefficient evaluated from coincidence detection on the remote polarization sensors.

As can be seen from the remote detection of the entangled signal photons polarization their statistic is in direct connection with reflectance changes. The calibrated polarization anisotropy gives an absolute temperature (Eq. 5). In the case described here we just take a linear calibration factor with initial and maximal temperature:

$$T(t) = T_0 + \frac{T_{max} - T_0}{A_{hot} - A_{cold}} (A(t) - A_{cold}) \tag{6}$$

The method described here of remote temperature evaluation has a temporary inertia according to average window width, but in principle it is directly connected to the frequency of pulse generation. For the window size near 40 ns and 1 photon event step it gives a time shift of temperature maximum near 200 ns (Fig. 7). For the shorter window and the fixed decay time of single photon detection the signal becomes more unstable.

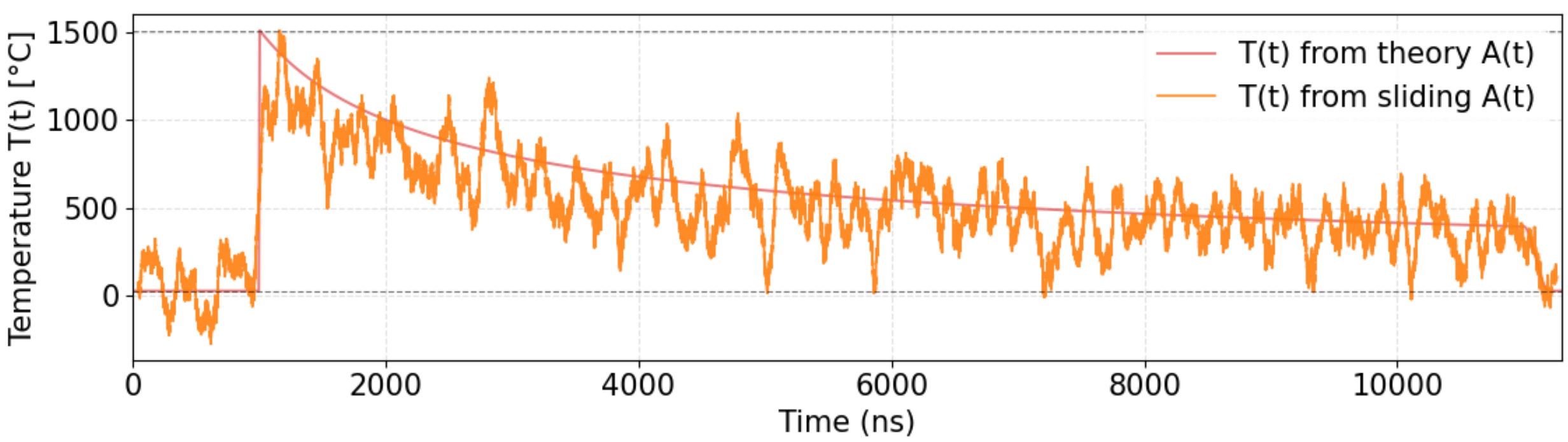

Fig. 7. Temperature evaluation by calibration of polarization anisotropy and averaging window duration equal to 40 ns.

The achieved results have a representative tendency of temperature changes, but the root mean square amplitude of noise reaches up to 200 °C. In that situation the signal to noise ratio is low, down to 10 dB and it is difficult to extract the clear dynamics of temperature changes. If the $H$-photons absorption increases, the situation becomes worse - the measured data will be represented as a hundred of thousands long vector of zeros with randomly distributed "1". In this situation higher physical affection of material brings higher noise and extraction of positive signals becomes more difficult. For the correct evaluation of temperature changes from the binary measurement data, it is preferable to use an alternative data-processing method. In the next section, we describe an algorithm specifically designed to extract hidden information from long one-dimensional data vectors.

## 3. Neuronetwork processing of optical anisotropy binary vector

A fundamental dilemma arises from the discrete character of the measurement data in entangled photon polarization anisotropy, which are represented as H- and V-states in the form of a long binary vector [01101…00101] (Fig. 1). When the optical anisotropy is evaluated as the local density of a binary vector, the size of the evaluation window becomes critical: a longer window decreases the measurement resolution, whereas a shorter window increases noise due to fluctuations in the density arising from the binary switching between '1' and '0' (Fig. 7.).

In the case of one-dimensional binary vector analysis, the application of a neural network is more suitable, particularly when the information about optical anisotropy is encoded indirectly and cannot be easily extracted using analytical methods. It should be emphasized that the primary objective of this study is not the development or deep optimization of machine-learning (ML) models, but rather the evaluation of whether a standard ML approach can provide a more robust alternative to the classical sliding-window method for temperature reconstruction under realistic noise conditions. In this context, the ML model is treated as a representative data-driven estimator and is directly compared with the analytical sliding-window approach using identical input data.

The uncertainty of both methods is therefore assessed consistently through the statistical spread of reconstruction errors across the dataset, without introducing additional model-specific uncertainty estimation techniques. While more advanced analyses—such as pointwise confidence intervals, bootstrap-based uncertainty estimation, or alternative kernel-based reconstruction methods (e.g., Gaussian-kernel smoothing)—could be considered, they fall outside the scope of the present work. The focus of this study is thus a direct, application-oriented comparison demonstrating the relative robustness of the two approaches in the presence of accidental coincidence noise.

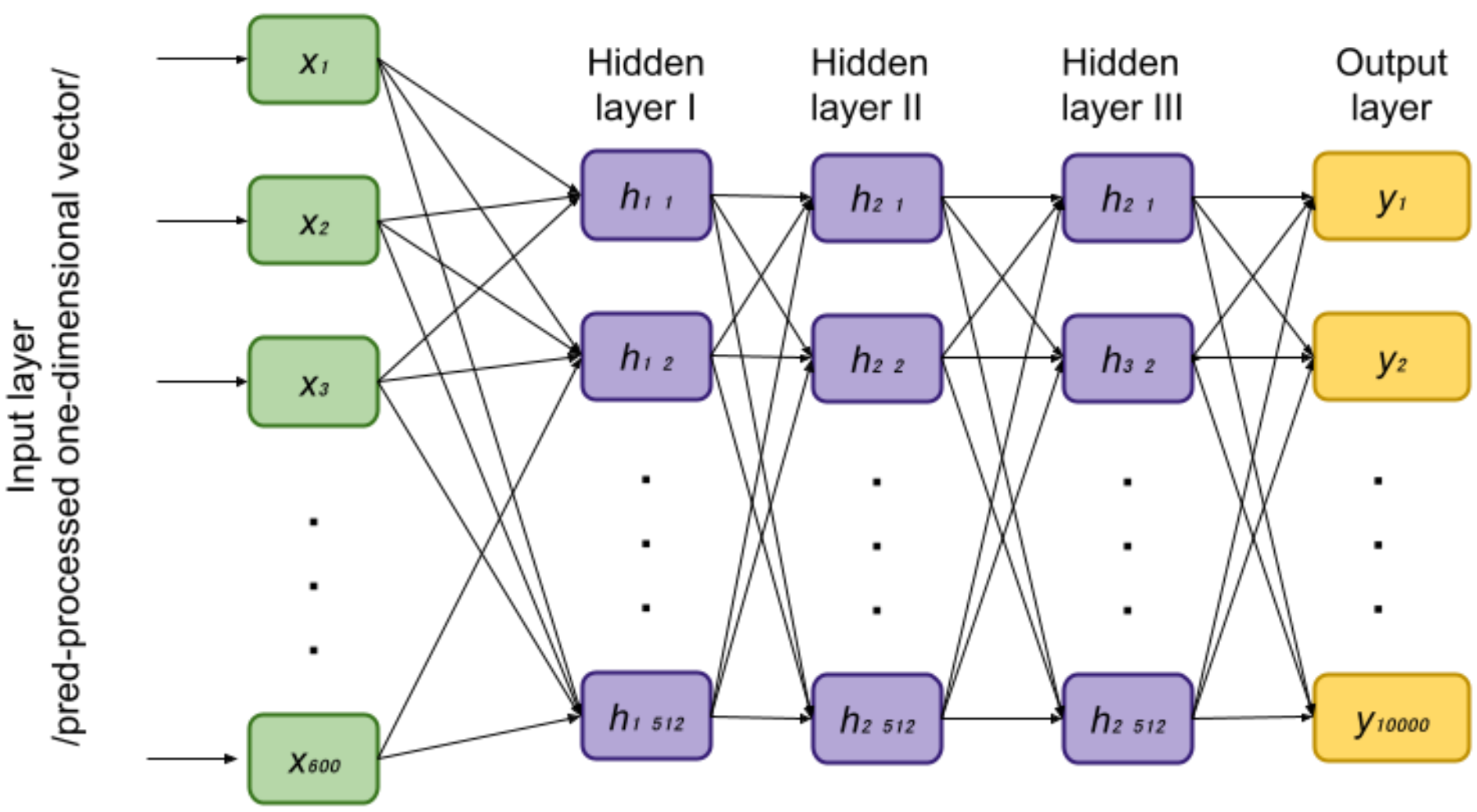

Fig. 8. Structure of neural network for temperature evaluation with three hidden layers.

To address this straightforward temperature-evaluation problem, a multilayer perceptron (MLP) is commonly employed. It is a fundamental type of artificial neural network (ANN) and has been widely used in infrared (IR) data processing over recent years [29]–[31]. For the binary vector processing a typical three hidden layers scheme will be used - it is light weight and efficient enough for temperature dynamic reconstruction (Fig. 8). The raw binary bitstream is converted into a fixed-length input vector suitable for neural-network regression:

$$bin_k = \{b_i |\ i \in [i_k, i_{k+1}]\},\ k = 1 \dots N \tag{7}$$

where $bin_k$ is equal-sized segments of the raw binary vector, $b_i$ - is entangled photon polarization state represented in binary form, $N$ is the full number of used bins. For the each bin the $H$-fraction is computed as statistical estimate of local polarization anisotropy:

$$x_k = \frac{1}{|bin_k|} \sum_{b_i \in bin_k} b_i, \tag{8}$$

The analytical temperature evaluation employed the same statistical estimation formula, but utilized a 40 ns window size, which is two times wider in comparison to the 18 ns window in the MLP model. Despite this adjustment, the analytical method yielded a high root mean square amplitude, about 20% of the positive signal amplitude.

In contrast, in the MLP-based method of temperature evaluation the noise suppression is achieved by extraction of hidden information. It can be explained in the way that each combination of “1” and “0” corresponds to its original temperature evaluation, with high sensitivity to local density in a bin. As a result, the MLP achieves an order-of-magnitude lower RMS noise ~20 °C and a much higher signal-to-noise ratio (SNR) near 26 dB (Fig. 9).

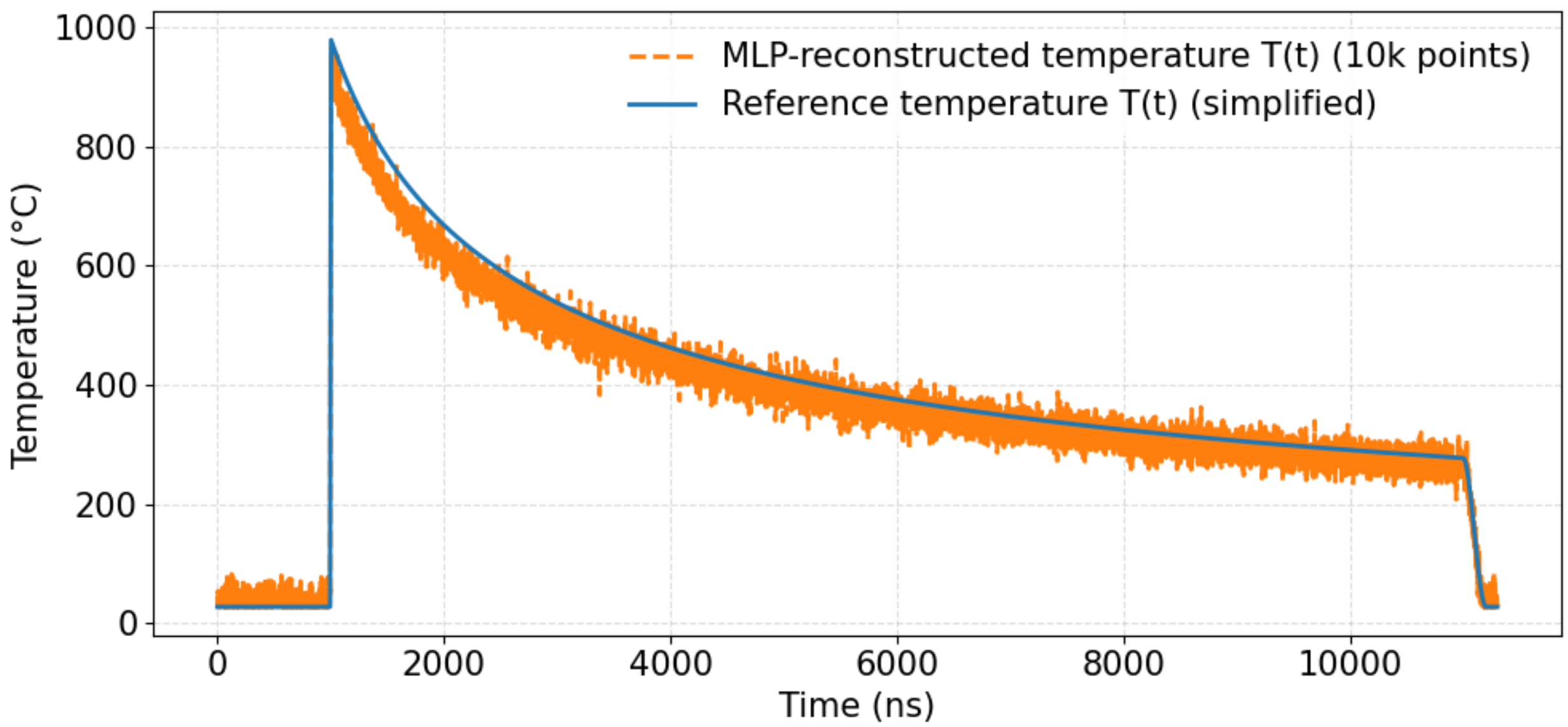

Fig. 9. MLP-reconstructed temperature (orange line) compared with the reference temperature (solid blue line).

In the model presented here, the input is a vector consisting of six hundred statistical estimation bins (Eq. 8). It decreases the neural network size for applications on regular CPU cores. The full network architecture can be described mathematically as consequence function:

$$\mathbf{h}^{(1)} = \mathrm{ReLU}\left(W^{(1)}\mathbf{x} + \mathbf{b}^{(1)}\right) ,$$
$$\mathbf{h}^{(2)} = \mathrm{ReLU}\left(W^{(2)}\mathbf{h}^{(1)} + \mathbf{b}^{(2)}\right) , \tag{9}$$

$$\mathbf{h}^{(3)} = \mathrm{ReLU}\left(W^{(3)}\mathbf{h}^{(2)} + \mathbf{b}^{(3)}\right),$$

$$y = W^{(4)}\mathbf{h}^{(3)} + \mathbf{b}^{(4)}.$$

where $\mathbf{h}^{(f)}$ - is the hidden-layer activation vector, ReLU is the nonlinear activation for collapsing model into a single linear transform, $W^{(f)}$ and $b^{(f)}$ are the trainable weights and biases of layer $f$, and $y$ is the point reconstructed temperature curve. The MLP model used here contains approximately ~7.8 million trainable parameters. It is a middle-sized model with the capacity of the non-convolutional architecture suited to converting coarse-grained photon statistics into temperature evaluation with 1 ns detailing (Fig. 9, orange line).

To evaluate the performance of the proposed neural-network-based temperature reconstruction, it was directly compared with a classical sliding-window method applied to the same binary coincidence data. The sliding-window approach estimates temperature locally by averaging coincidence statistics within a finite temporal window, whereas the ML model performs a global mapping from the full input vector to the temperature evolution, implicitly accounting for correlations and noise patterns in the data. The reconstruction accuracy of both methods was quantified using the root-mean-square error (RMSE) [32], defined as:

$$RMSE = \sqrt{\frac{1}{N}\sum_{i=1}^{N}\left(T_i^{rec}(t_i) - T_i^{true}\right)^2}$$

where $T_i^{rec}$ and $T_i^{true}$ denote the reconstructed and reference temperatures, respectively, and $N$ is the number of time points. This metric provides a direct measure of the temperature evaluation error in physical units (°C) and enables a consistent comparison between the two reconstruction approaches.

## 4. Discussion

The presented results have shown that entangled photons can be used for fast detection of surface temperature changes. In the simulation, the periodicity of the entangled pairs generation of around 50 ps was used. According to the described principle, the temperature changes during rapid surface heating were evaluated from the ratio between $H$- and $V$-state counts in the remote signal arm, while the idler photons were used as a coincidence trigger (Eq. 5). The evaluation of absolute temperature based on experimental data with pulsed ellipsometry measurements [14] gave realistic results. In principle the calibration of the remote temperature evaluation with entanglement photons can be done by preliminary measurements on a hot plate of the used material. Due to the quantised character of the entanglement photon detection it is impossible to evaluate polarization anisotropy just by one event and it is needed to have some statistical window. The width of the used window was equal to 40 ns and it gives the time shift of the temperature rise of around 200 ns and SNR around 10 dB. The statistical character of such measurement imposes a limitation on the time resolution and binary nature of $H$- and $V$-state output vectors. The application of neural network algorithms to this type of data has demonstrated improved precision. For a model with 600 statistical estimation bins and more than seven million trainable parameters, the temporal shift of the temperature rise is approximately 8 ns, while the signal-to-noise ratio exceeds 26 dB. Such a difference of the binary data processing is explained by the basic principle of the discrete measurements. This can be explained as shown in Fig. 6, where absorption of $p$-polarized photons is higher, but it also results in increased noise. It is just by more rarely appearing of “1” state detection in comparison to untriggered photons. As a result, the measured data contain a more randomly distributed pattern of “1” bits within an array of “0” bits. That is why the analytical method of such data processing has not worked properly and MLP method becomes beneficial.

It should be emphasized that classical temperature-detection techniques offer several advantages. These include direct access to the complex refractive index $n + ik$ through mature modeling approaches with sub-millidegree precision in ellipsometry, and reflection-free temperature measurement in pyrometry. But the entangled photons polarization anisotropy detection has great advantage in background rejection via coincidences. Only idler detections that arrive within the coincidence window of an entangled signal photon are counted. This suppresses the stray light, furnace glow, and detector dark counts far better than classical ellipsometry or pyrometry (which integrate all photons hitting the detector). The next feature is the heralded detection with known arrival times. The signal photon “announces” the idler and picosecond (or sub-ns) coincidence gates can be used, even when single-photon pulses are weak. The next benefit of the entangled photons application lies in the wavelength optimization for the metal reflectance for low-noise with high-QE detectors. Classical pyrometry is constrained by thermal-emission spectral bands, whereas ellipsometry requires an intense probe beam and high-sensitivity detectors at the probe wavelength. The discrete character of the entanglement photons polarization anisotropy detection has ultralow probe flux (non-perturbative). It gives higher SNR above backgrounds, orders-of-magnitude fewer photons than classical ellipsometry. It is especially important when additional heating or ablation must be avoided. Another distinctive feature of the H- to V-state ratio principle is its immunity to emissivity drift. In contrast to pyrometry, in which temperature evaluation is sensitive to unknown or varying emissivity and view factors, the proposed method is largely immune to these effects. Temperature evaluation with polarization anisotropy has the shot-noise suppression potential. With the correlated pairs and proper statistics, the classical shot-noise scaling is beaten for the same detected flux with improving precision per detected photon. It should be noted that the measured data are represented as a long binary vector, for which classical analysis methods are not effective in extracting the signal. On the other hand, the application of neural network analysis is able to extract implicit information for correct evaluation of thermal process dynamics. This is clearly seen by comparing the results of the sliding-window algorithm (Fig. 7) with those of the MLP data-processing model (Fig. 9).

The reconstruction accuracy of the sliding-window method and the ML model was evaluated over 150 generated datasets with controlled accidental coincidence noise. The accidental contribution was varied up to approximately 95% of all detected coincidence events. The resulting temperature evaluation error, quantified by RMSE (Section 3), is summarized in Fig. 10 as a function of the accidental fraction. Under near-ideal conditions, both methods reproduce the general temperature dynamics; however, the sliding-window approach exhibits significantly larger errors (RMSE ≈ 320 °C) compared to the ML model (RMSE ≈ 28 °C). As the accidental coincidence fraction increases, the reconstruction accuracy deteriorates for both methods, but the degradation is considerably stronger for the sliding-window approach. At intermediate accidental levels, its RMSE increases to approximately 600 °C, whereas the ML model maintains substantially lower errors of about 300 °C. At the highest tested noise level, corresponding to approximately 95% accidental coincidences (about 20 times more accidental than true coincidence events), the sliding-window reconstruction exhibits both high error and large variability between datasets. In contrast, the ML-based approach remains comparatively robust, maintaining lower RMSE values and a much smaller spread of results. The behavior observed in Fig. 10 indicates that reconstruction based on local coincidence statistics is highly sensitive to accidental coincidences, whereas the neural-network model is better able to account for noise and preserve a physically meaningful temperature evolution over a broad range of measurement conditions.

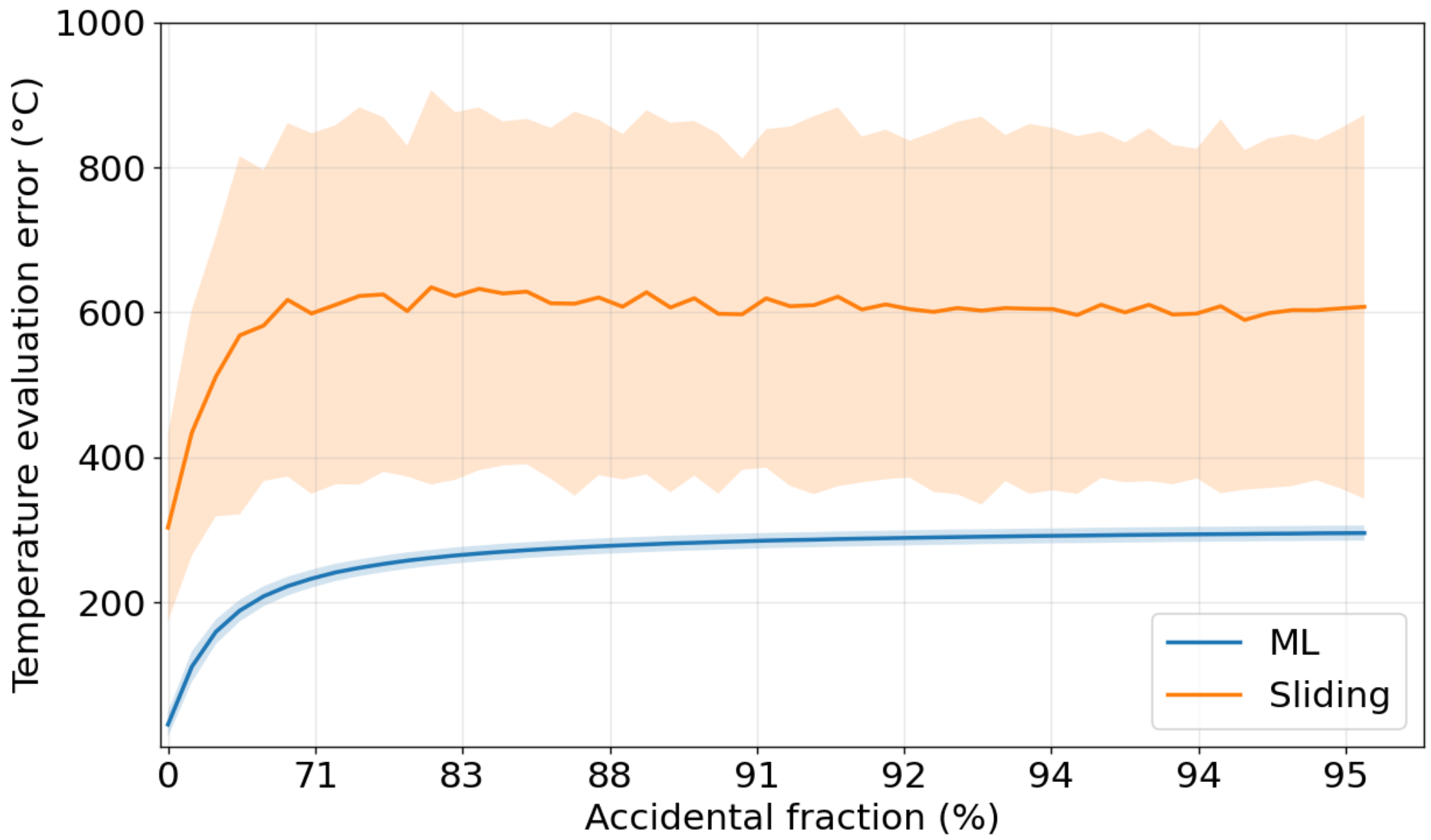

Fig. 10. Temperature evaluation error (RMSE) as a function of accidental coincidence rate (false pairs per pulse): blue - for the ML-based and orange - sliding-window reconstruction methods. The shaded regions indicate the standard deviation over the dataset.

The middle-sized ANN used in this work can be further improved by training with higher-resolution data or by switching to raw-data processing instead of statistical estimation bins. However, in this case, the neural network would need to be adapted for GPU acceleration and extended to a deeper architecture. It makes sense in the case of real applications where the output values are predefined for a narrow task. The goal of this work is to show an alternative method for temperature measurement based on quantum effects with binary data format processing.

Another possible approach to fast temperature evaluation using entangled photons is based on a hybrid methodology combined with discrete Fourier transform (DFT)-based data processing. In this scheme, coincidence detection is employed as a triggering mechanism together with measurements of the reflected-light intensity and fast pyrometry along a single optical path. The binary character of the measurement provides additional benefits beyond the evaluation of coincidence events alone, enabling the analysis of background-related detection statistics as well. In particular, non-coincidence events, in which an idler-side detection occurs without a corresponding entangled partner in the signal arm, can indicate additional thermal emission from the laser-heated surface. Conversely, coincidence-lost events (where a signal photon is detected without a registered idler photon) may arise from increased absorption, scattering, or surface deformation on the idler side. The temporal evolution of the rates of such 'lost' events can therefore serve as an additional indicator of laser-induced modifications of the processed surface. Incorporating these auxiliary detection statistics enhances the overall dataset and improves the precision of optical temperature measurements. Such a hybrid approach can further extend the understanding of fast physical processes initiated by the interaction of ultrashort laser pulses with matter.

## 5. Conclusion

In this article, polarization anisotropy of entangled photons is introduced as a promising approach for fast temperature measurement with remote detection. The main advantages include suppression of disturbing effects such as changes in emissivity of the laser-processed material and plasma emission, as well as potential real-time response in the nanosecond time range. The method was evaluated on the modeling of ultrashort laser pulse interaction with stainless steel. The principle of temperature measurement based on remote polarization detection, triggered by reflected idler photon events, was demonstrated. Such separation of the triggering and polarization measurement stages reduces the requirements for local measurement techniques, as only a single-photon bucket detector is needed in the laser interaction zone for idler photon registration. A key advantage of the proposed method is that temperature is derived from the coincidence rate rather than from the absolute signal intensity. This enables suppression of parasitic effects associated with idler reflection, such as surface deformation, scattering from roughness, or emissivity variations of the laser-irradiated material.

A quantum circuit model was employed to simulate the measurement principle. First, a Hadamard gate was applied to the signal qubit to prepare a superposition state, followed by a CNOT gate to generate entanglement. Coincidence detection was then realized by measuring both qubits. In the used scheme, optical anisotropy is detected not by the intensity of the reflected signal, but by the relative amount of H- and V-state remotely detected signal photons. The simulation indicates that analytical recognition of temperature by the sliding-window method has inherent limitations due to the binary nature of single-photon measurements. The selected short window size (40 ns), required for good temporal resolution, leads to a relatively low signal-to-noise ratio (~10 dB).

An alternative method of results processing was investigated, based on a regression neural network applied to entangled photon sequence analysis. This approach improves the effective SNR (26 dB versus 10 dB) and provides more stable reconstruction under noisy conditions. Although the MLP-based reconstruction exhibits reduced temporal resolution, it enables faster detection of temperature rise in the considered scenario. The robustness of the ML model against accidental coincidences was evaluated using RMSE as a function of noise level, demonstrating significantly higher stability compared to the sliding-window method. While the classical approach becomes unstable at high noise levels (up to 12% accidental pairs), the ML model maintains controlled error and lower variance. These results indicate that data-driven reconstruction can mitigate the impact of noise in entangled-photon-based measurements. A more detailed uncertainty analysis, including confidence intervals and kernel-based methods, may be considered in future work. The presented results suggest that neural-network-based processing can extend the applicability of entangled-photon temperature measurements toward high-speed regimes. Further development may include the use of weak measurement schemes [33], where indirect measurements combined with neural-network-based reconstruction could provide improved sensitivity and resolution.

**Acknowledgement**

The work was supported by the Ministry of Education, Youth and Sports of the Czech Republic (Programme Johannes Amos Commenius, co-funded by EU, call Excellent Research, project MEBioSys, No. CZ.02.01.01/00/22_008/0004634, and project QM4ST, No. CZ.02.01.01/00/22_008/0004572).